\documentclass[conference]{IEEEtran}
\IEEEoverridecommandlockouts
\usepackage{cite}
\usepackage{amsmath,amssymb,amsfonts}
\usepackage{algorithmic}
\usepackage{graphicx}
\usepackage{textcomp}
\usepackage{xcolor}
\usepackage{enumitem}
\usepackage[inkscapelatex=false]{svg}
\def\BibTeX{{\rm B\kern-.05em{\sc i\kern-.025em b}\kern-.08em
    T\kern-.1667em\lower.7ex\hbox{E}\kern-.125emX}}

\usepackage{tikz}

\newcommand\copyrighttext{%
  \footnotesize \textcopyright \the\year{} IEEE. Personal use of this material is permitted. Permission from IEEE must be obtained for all other uses, including reprinting/republishing this material for advertising or promotional purposes, collecting new collected works for resale or redistribution to servers or lists, or reuse of any copyrighted component of this work in other works.}

\newcommand\copyrightnotice{%
\begin{tikzpicture}[remember picture,overlay]
\node[anchor=south,yshift=10pt] at (current page.south) {%
\begin{minipage}{\textwidth}
\center \copyrighttext
\end{minipage}};
\end{tikzpicture}%
}

\newcommand\acceptedtext{%
  \footnotesize This article has been accepted for publication in the proceedings of 2025 IEEE 12th International Workshop on Metrology for AeroSpace (MetroAeroSpace),
  but has not been fully edited. Content may change prior to final publication. \\
  Citation information: DOI 10.1109/MetroAeroSpace64938.2025.11114598, 2025 IEEE 12th International Workshop on Metrology for AeroSpace (MetroAeroSpace).}
  
\newcommand\acceptednotice{%
\begin{tikzpicture}[remember picture,overlay]
\node[anchor=north,yshift=-2pt] at (current page.north) {%
\begin{minipage}{\textwidth}
\center \acceptedtext
\end{minipage}};
\end{tikzpicture}%
}

\begin{document}

\title{A Methodological Framework for Positioning of
Wireless Sensors in New Generation Launchers\\
\thanks{This research was funded by ASI (Italian Space Agency), Agenzia
Spaziale Italiana, within the project “Avionics systems for launcher of the
future generation”, grant number N. 2022-15-I.0, CUP F83D22001250005, CIG 8421342742.
The research activities presented in this paper fall within the field of interest of the IEEE AESS technical panel on Glue Technologies for Space Systems.}
}

\author{
\IEEEauthorblockN{
Ivan Iudice\IEEEauthorrefmark{1},
Domenico Pascarella\IEEEauthorrefmark{1},
Sonia Zappia\IEEEauthorrefmark{1},
Giovanni Cuciniello\IEEEauthorrefmark{2},\\
Hernan M. R. Giannetta\IEEEauthorrefmark{3},
Marta Albano\IEEEauthorrefmark{4}
and
Enrico Cavallini\IEEEauthorrefmark{4}
}
\IEEEauthorblockA{
\IEEEauthorrefmark{1}%
Security Unit
and
\IEEEauthorrefmark{2}%
Avionic Systems Unit,
Italian Aerospace Research Centre (CIRA), Capua (CE), Italy
}
\IEEEauthorblockA{
\IEEEauthorrefmark{3}%
Research and Development Department, SpaceLab, Rome, Italy
}
\IEEEauthorblockA{
\IEEEauthorrefmark{4}%
ASI (Italian Space Agency), Rome, Italy
}
\IEEEauthorblockA{
Email: [i.iudice, d.pascarella, s.zappia, g.cuciniello]@cira.it,\\
HernanMaximilianoRoque.Giannetta@spacelabcompany.it,
[marta.albano, enrico.cavallini]@asi.it
}
}

\maketitle
\copyrightnotice
\acceptednotice

\begin{abstract}
In wireless sensor networks for reusable launchers, the electromagnetic characterization and electromagnetic compatibility analyses are relevant due to the reference operational scenario, which implies a complex, and sometimes dynamic, electromagnetic environment. This work proposes a methodological framework for the design of the network and for the analysis of the related electromagnetic environment within the stages of a given launcher. Based on the preliminary positioning of the network nodes, the framework prescribes a workflow and the related toolset for determining the optimal network topology focusing on the weights, the operation of the transceivers, and the overall radiated power. The optimal network configuration is simulated by using computational electromagnetics strategies in order to assess the electromagnetic environment induced by the sensor network itself. The paper provides some results concerning a case study for a specific launcher.
\end{abstract}


\section{Introduction}
Many stakeholders are evaluating the use of wireless technology inside spacecrafts due to its advantages, like weight reduction, simplified integration and unrestricted instrument mobility \cite{Fusco2024}. Such evaluation requires electromagnetic analyses to assess the electromagnetic interactions of the wireless system within the spacecraft.
In wireless sensor networks (WSNs) for reusable launchers, the \emph{electromagnetic characterization} (ELM) and \emph{electromagnetic compatibility} (EMC) analyses are relevant due to the reference operational scenario, which implies a complex, and sometimes dynamic, electromagnetic environment for the WSN system.

In detail, such analyses generally contribute to reach the following main goals:
\begin{itemize}
    \item Interference Prevention and Reliable Communication in the WSN – Electromagnetic emissions of the WSNs shall not interfere with other critical onboard systems (avionics, propulsion control, flight computers, etc.), which rely on interference-free communication. In addition, the wireless sensor network system shall ensure reliable communication for real-time data transmission, even in challenging conditions due to the presence of interferences coming from other systems. In case of compromise of the communication reliability, critical data could be corrupted, leading to safety risks or mission failures. This especially holds for critical sensors, e.g., in charge of the real-time monitoring of the status of the vehicle.
    \item Electromagnetic-Emission Minimization in the WSN – Exceeding emission limits could cause interference with other systems. The optimization of spectrum usage contributes to prevent interference in order to ensure reliable communications. On the other hand, it improves the overall launcher performance by minimizing the need for complex shielding or filtering solutions and limiting SWaP requirements.
\end{itemize}

To this end, the network topology design represents a critical point of the sensor network deployment; furthermore, the electromagnetic scenario of the WSN system shall be assessed by means of electromagnetic analyses and simulations of the sensor network implementing the given topology.

This work proposes a methodological framework for the design of the WSN and for the analysis of the related electromagnetic environment within the stages of a given launcher. Based on the preliminary positioning of the nodes constituting the WSN, the framework prescribes a workflow and the related toolset for determining the optimal network topology focusing on the weights, the operation of the Radio Frequency (RF) transceivers, and the overall radiated power. The optimal network configuration is simulated by using computational electromagnetics (CEM) strategies in order to assess the electromagnetic environment induced by the sensor network itself.

\section{Background}

Theoretically, for designing the sensor network topology, CEM tools can aid for deterministically determining the signal characteristics (e.g., received power, delay spread, power spectrum) and, thus, optimizing the desired objective function.
In practice, the actual representation of the electromagnetic environment is not achievable, since:
(i) it would need the perfect characterization of the obstacles (i.e., objects and/or structures) surrounding the RF devices, in terms of materials and geometry (i.e., CAD models with an accuracy at least comparable with the wavelength); small displacement of RF devices and/or obstacles, however at the wavelength scale, can lead to significant variation in the simulation results;
(ii) often it is not possible to obtain the needed electromagnetic characteristic details of all of the devices composing the launcher electronics;
(iii) the exhaustive electromagnetic simulation can be too complex to be carried out by using general purpose hardware architectures, especially when full-wave solvers need to be used (e.g., due to near-field effects).

The most effective approach for designing wireless networks is based on hybrid statistical-empirical models \cite{Gol2005}. Specifically, statistical methods are used for modelling specific impairments (e.g., shadowing, fading), on the other hand, empirical data are used for determining the parameters of the statistical models relating to specific scenarios. Such approach is more general, and it is less sensitive to small variations in the context.

In \cite{Garbarino2024}, the authors analyse the design of a wireless sensor array system for a spacecraft inflatable habitat. In detail, a wireless structural health monitoring system is designed and simulated using the shooting and bouncing rays numerical electromagnetic technique. In \cite{Wang2022}, the authors investigate the calculation of electromagnetic interference margin of spacecraft wireless system by simulating the isolation degree of the transceiver antenna. In \cite{Yan2023}, the authors analyse the electromagnetic interferences of wireless devices inside the airproof cabin of manned spacecraft, by using a structure model of one manned spaceship and by setting the related electromagnetic simulation model. Lastly, other works provides a detailed survey about wireless communication for space applications \cite{Castrillo2024}, also considering the specific case of WSN \cite{Yukun2022}.

To the best of our knowledge, this is the first work to propose a methodological framework for determining the optimal sensor network topology and analysing the electromagnetic environment concerning WSNs inside launchers.

\section{Methodological framework}

This section describes the proposed methodological framework, meant as a structured set of workflow, techniques and tools for determining the optimal sensor network topology and analysing the electromagnetic environment of a WSN within a new generation launcher. In detail, the next subsections elaborate on the workflow and the toolset of the proposed framework.

\subsection{Workflow}\label{sec_workflow}
\begin{figure*}[htbp]
\centerline{\includegraphics[width=0.7\linewidth]{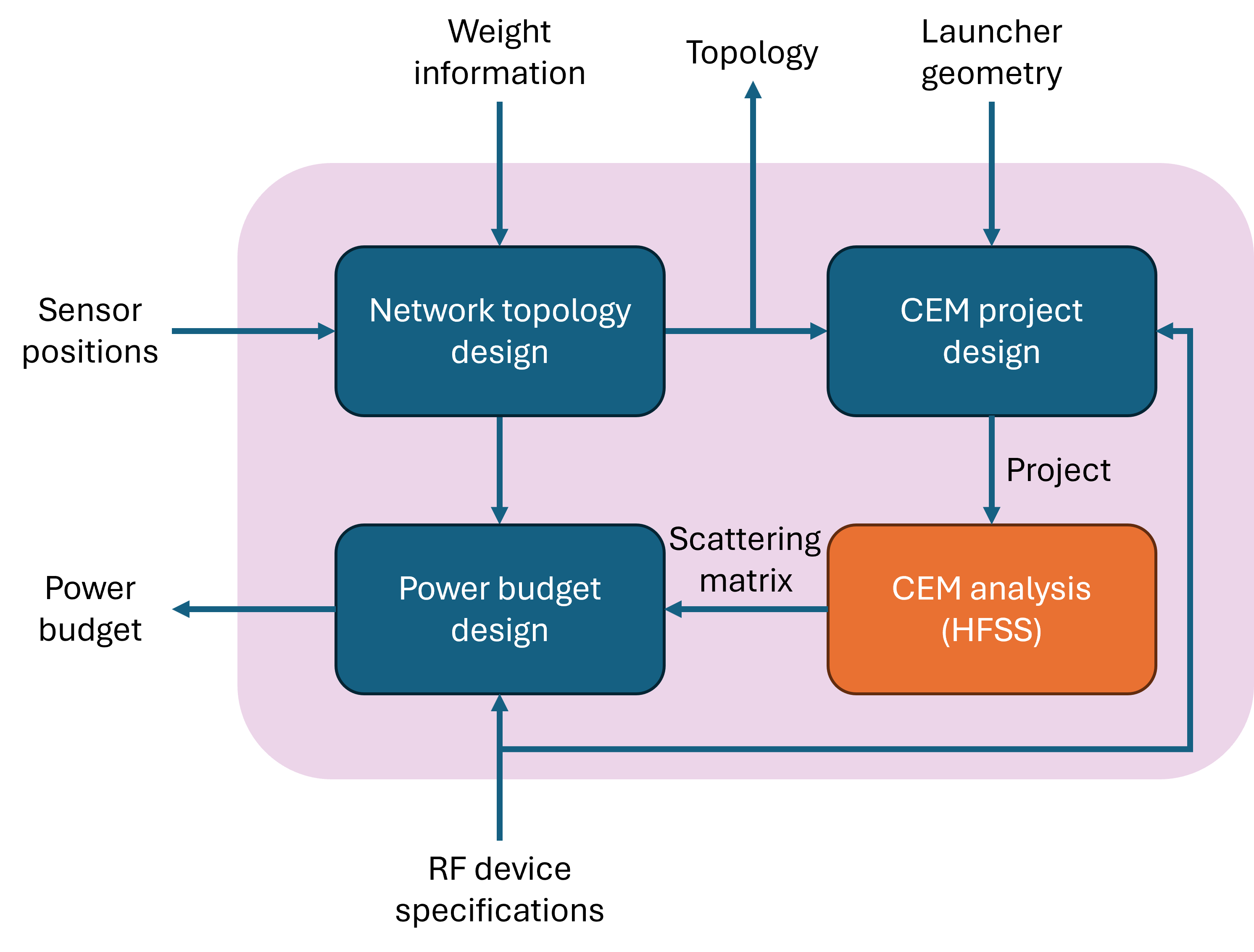}}
\caption{Workflow diagram.}
\label{fig:workflow}
\end{figure*}

Fig.~\ref{fig:workflow} reports the pictorial representation of the proposed workflow. This may be applied to each single stage of the launcher or arbitrary segments of it. Specifically, it is possible to identify segments that can be assumed electromagnetically or functionally isolated. As an example, subsets of sensors belonging to the same stage can be considered electromagnetically isolated by the attenuation provided by the propellent; on the other hand, sensors belonging to different stages (connected to different HUBs) can share some volumes even though they are functionally isolated due to the medium access control (MAC) provided by the protocol implemented by the wireless sensor network (e.g., 802.11ah). For the explanation of the colors of the block in Fig.~\ref{fig:workflow}, see subsection \ref{sec_toolset}.

The workflow requires the following inputs: the launcher geometry; the positions of the sensors installed into the launcher; weight information in terms of the unit weight of each of the RF devices (i.e., sensor RF kit, HUB) and the weight per meter of the cable used for wired communication connections. The workflow delivers the following outputs: the optimal WSN topology; the optimal power budget comprising the power to be used at each of the RF transmitters, and the overall power emitted.

The next subsections report a detailed description of the techniques applied within the workflow steps.

\subsection{Network Topology Design}
In general, providing a sensor RF kit for each of the sensors would lead to an overall weight greater than the all-wired solution. So, it is important to use fewer sensor RF kits acting as multiplexers for multiple sensor data sources. Unfortunately, without specific information on the cable paths or on the harness constraints needed for connecting wired sensors with the sensor RF kits, the linear distance is the only parameter that can be considered for evaluating a lower bound for weights.

Also regarding the wireless connections, general-purpose CEM solvers are not suitable for conducting extensive optimization procedures that would require fast solutions of Maxwell’s equations exploiting application-specific characteristics. Thus, in our framework, network topology design may be carried out by neglecting all multipath propagation, shadowing, and near-field effects.

Specifically, considering only the path loss, it is well known that the received power can be evaluated using the Friis transmission formula \cite{Gol2005},

\begin{equation}
P_{\text{RX}}=\frac{P_{\text{TX}}G_{\text{TX}}G_{\text{RX}}}{L_{0}}\left( \frac{d_{0}}{D} \right)^{\gamma},\label{eq_Friis}
\end{equation}

from which it is apparent that the linear distance $D$ between $TX$ and $RX$ is the only parameter depending on the network topology to be optimized.

For the above reasons, the proposed topology design approach is based on:
\begin{enumerate}
    \item clustering the sensors depending on the mutual distance and the number of sensor RF kits to be used;
    \item placing each sensor RF kit in correspondence with the median point of the relative sensor cluster;
    \item clustering the sensor RF kits depending on the mutual distance and the number of HUBs to be used;
    \item placing each HUB in correspondence of the centroid of the relative sensor RF kit cluster.
\end{enumerate}
Naming $N_{s}$, $N_{\text{RF}}$, and $N_{\text{HUB}}$ the number of sensors, sensor RF kits, and HUBs, respectively, the inequality $N_{\text{HUB}}<N_{\text{RF}} \leq N_{s}$ must hold.

For clustering, the K-means method \cite{Goodfellow2016} is suggested. Such algorithm is very fast (one of the fastest clustering algorithms available), but it can fall in local minima, so, it can be restarted several times and the best solution can be chosen.

Fixed $N_{s}$, different optimization solutions are obtained for different values of $N_{\text{HUB}}$ and $N_{\text{RF}}$. For each of the found solutions, the overall weight can be evaluated, finally, the network topology leading to the minimum weight can be chosen. Note that when a HUB is requested to serve only one sensor RF kit, the HUB is dropped and the sensor RF kit will be cabled for working as a simple multiplexer of multiple sensor data sources. The number of clustering needed for the overall parametric optimization is $N_{\text{RF}}(N_{\text{HUB}}+1)$.

\textcolor{black}{Finally, the proposed approach provides a sub-optimal topology due to the considered assumptions and design choices, i.e., (i) no information on the cable paths and on the harness constraints, (ii) neglecting multipath propagation, shadowing, and near-field effects, (iii) adoption of the K-means method for clustering the locations of RF kits.}

\subsection{CEM Project Design and Analysis}

A simplified model of the electromagnetic environment is considered for CEM simulations. Specifically, in order to reduce the computational complexity of the CEM evaluations, a Method of Moments (MoM) based solver \cite{Davidson2010} is considered for the framework.
MoM based solvers exhibit great efficiency when modelling perfectly conducting objects and sheets; the main advantage consists on the needing to mesh and evaluate only the surface currents without considering the surrounding vacuum.
The aim of the CEM is to compute the scattering matrix of all the RF devices (i.e., sensor RF kits and HUBs) to be used for evaluating the received power, also considering multipath fading due to reflections occurring at the interior walls of the cylinder.

\subsection{Power Budget Design}

When the scattering matrix is known, it is possible to evaluate how much power each transmitter has to employ to reach the sensitivity of all of the receivers. Specifically,
\begin{equation}
P_{\text{RX},j}^{\left(s\right)}=S_{ij}^{\left(s\right)} P_{\text{TX},i}^{\left(s\right)},\label{eq_received_power_j}
\end{equation}
where $P_{\text{RX},j}^{\left(s\right)}$ represents the received power at the $j$-th RF device belonging to the segment $s$, when the $i$-th RF device of the segment $s$ is transmitting the power $P_{\text{TX},i}^{\left(s\right)}$, and $S_{ij}^{\left(s\right)}$ is the $\left(i,j\right)$-th element of the scattering matrix. Determining the power budget for the $i$-th transmitting RF device consists in finding,
\begin{equation}
\overline{P}_{\text{TX},i}^{\left(s\right)} = \frac{P_{\text{min}}}{\underset{j}{\min} S_{ij}^{\left(s\right)}},
\label{eq_power_budget_i}
\end{equation}
where $P_{\text{min}}$ represents the sensitivity of all of the RF devices%
\footnote{The evaluation above represents a worst case scenario. As an example, when time-division multiple access (TDMA) is used, the transmitting HUB can adapt the instantaneous transmission power in order to further reduce the average power emission.}.


The overall power emitted throughout all of the segments can be readily evaluated by,

\begin{equation}
\overline{P}_{tot}\vert_{\text{dBm}} = 10 \log_{10} \left( \underset{s}{\sum}\underset{i}{\sum} \overline{P}_{TX,i}^{\left(s\right)}\right).\label{eq_total_emitted_power}
\end{equation}

Note that \eqref{eq_total_emitted_power} is the sum of the powers transmitted by all of the RF devices in the launcher, but the simplified electromagnetic environment
at hand is a sort of Faraday cage, only open at the bottom. The best way to evaluate the overall emission would be to simulate all of the sources, together with the overall structure of the launcher, in order to measure the power flow coming out from the bottom of the launcher. Such a simulation will be too expensive to be carried out using general purpose full-wave solvers, e.g., MoM; on the other hand, using general purpose asymptotic solvers may lead to significantly incorrect results due to the neglecting of near-field effects. Thus, once again, \eqref{eq_total_emitted_power} is given as worst-case reference.

\subsection{Toolset}\label{sec_toolset}

The workflow described in subsection \ref{sec_workflow} shall be implemented by a specific toolset for the purposes of the reference analysis. However, such toolset shall be also customizable for the different cases, i.e., when it is needed to match new specific requirements and new input data are available.
In the current version of the proposed methodological framework, the blue blocks in Fig.~\ref{fig:workflow} have been expressly implemented in the Python programming language, using open-source libraries (e.g., NumPy, Pandas, SciPy, scikit-learn, PyAEDT), on the other hand, the orange block has been implemented using the proprietary software ANSYS Electronics Desktop (AEDT).

\section{Results}
\textcolor{black}{The proposed methodological framework has been applied to a specific launcher reference design assumed within the project} for determining its optimal WSN topology, focusing on the weights, the operation of the RF transceivers, and the overall radiated power. The optimal network configuration is simulated by using computational CEM strategies to assess the electromagnetic environment induced by the WSN itself.

\begin{figure}[htbp]
\centerline{\includegraphics[width=\columnwidth]{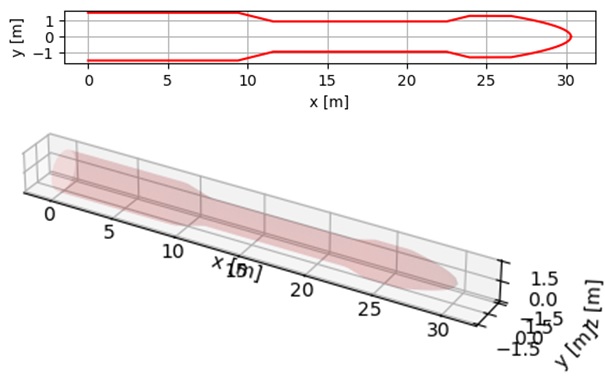}}
\caption{Synthesized launcher geometry.}
\label{fig:launcher_geometry}
\end{figure}

The following inputs have been used:
\begin{itemize}
    \item Launcher geometry - The launcher is approximated with a circular cross-section cylinder with variable radius, so that the only input information needed is the behaviour of the radius along the launcher axis. A piecewise linear variation of the radius has been assumed; only for the cap at the top of the launcher, a paraboloid shape has been considered. Fig.~\ref{fig:launcher_geometry} reports the synthesized launcher geometry.
    \item Sensor positions - A preliminary positioning of the WSN nodes has been specified by means of a comma separated values (CSV) file.
    \item Weight information - Table~\ref{tab:inputs_weights} reports the input weight information in terms of unit weight of each of the RF devices (i.e., sensor RF kit, HUB), and the weight per meter of the cable used for wired communication connections.
    \item RF device specifications - Table~\ref{tab:inputs_RF} reports the specifications of the RF devices composing the WSN.
\end{itemize}

\begin{table}[htbp]
\caption{Detailed weight information}
\begin{center}
\begin{tabular}{|c|c|c|}
\hline
\textbf{Item} & \textbf{Weight} & \textbf{Unit} \\
\hline
Sensor RF kit & 1.280 & kg \\
HUB & 1.880 & kg \\
Cable & 0.014 & kg/m \\
\hline
\end{tabular}
\label{tab:inputs_weights}
\end{center}
\end{table}

\begin{table}[htbp]
\caption{Detailed RF specifications}
\begin{center}
\begin{tabular}{|c|c|c|}
\hline
\textbf{Specification} & \textbf{Valuet} & \textbf{Unit} \\
\hline
Frequency band & 750-950 & MHz \\
Maximum transmit power & 17 & dBm \\
Maximum transmit gain & 30 & dB \\
Maximum antenna gain & 2 & dB \\
Receiver sensitivity & -109 & dBm \\
\hline
\end{tabular}
\label{tab:inputs_RF}
\end{center}
\end{table}

For the CEM project design, since no further details are often available, the launcher can be represented as a perfectly conducting circular cross-section cylinder with radius varying along the symmetry axis $x$, opened at the bottom, and closed at the top. Furthermore, for further reducing the computational complexity, surface currents are considered practically vanished after $10 \lambda_{\text{c}}$, thus, the launcher geometry has been partitioned in stages ranging between  $x_{\text{min}}-10\lambda_{\text{c}}$ and $x_{\text{MAX}}+10\lambda_{\text{c}}$, with $x_{\text{min}}$ and $x_{\text{MAX}}$ representing the minimum and the maximum $x$-coordinates of the RF devices belonging to the segment, respectively, and $\lambda_{\text{c}}$ representing the wavelength at the considered central frequency.

In order to obtain results as independent as possible from the shape of the field source, for each RF device an ideal point source of current oriented along $x$-axis has been taken as reference. Such an idealization has been implemented in CEM software as a very small dipole \cite{Harrington2001}.

Note that such scenario represents the most challenging at all, since when walls are not perfect electrical conductors, not all the power will be reflected with each bounce. Furthermore, when there are obstacles in the volume, the reflected waves are even more attenuated than the direct path due to the longer optical path.

%

The open-source based python scripts implementing some of the operations provided by the proposed workflow constitute the toolset of the framework and carry out the network topology design, the CEM project design and the power budget design. The only operation to be performed by a commercial software at the moment is represented by the CEM evaluation. The provided scripts can readily be adapted to support different CEM implementations (also open-source ones).

The network-topology script uses the K-means implementation provided by the scikit-learn Python library \cite{Pedregosa2011, Buitinck2013}, and the Pandas python library for data management \cite{Pandas2024}.
For the execution, the launcher has firstly been segmented in order to reduce the computational complexity of the algorithms, based on both the actual stages and the density of the sensors within them. Specifically, the first 3 stages have been cropped in the middle, leading to 8 isolated segments. Fig.~\ref{tab:num_sensors}
reports the number of sensors for each stage.

\begin{table}
\begin{center}
\caption{Number of sensors for each stage.}
\begin{tabular}{|c|c|}
\hline
Stage & $N_s$ \\
\hline
1a & 9 \\
1b & 26 \\
2a & 15 \\
2b & 8 \\
3a & 15 \\
3b & 7 \\
4 & 35 \\
5 & 44 \\
\hline
\end{tabular}
\end{center}
\label{tab:num_sensors}
\end{table}

For each of the aforementioned segments, the parametric optimization has been conducted considering the number of sensor RF kits ranging as $2\leq N_{\text{RF}}\leq N_{s}$, the number of HUBs ranging as $1 \leq N_{\text{HUB}}\leq3$, the number of run of K-means for each clustering procedure (with random initialization) equal to 1000, scattering matrices have been evaluated at the central frequency $f_{\text{c}}=850$ MHz.

In detail, the WSN topology optimization leads to $N_{\text{RF}}=2$ and $N_{\text{HUB}}=1$ for each segment. Thus, the found optimal topology requires the deployment of one single HUB and only two sensor RF kits, working as multiplexers for several sensor data sources, for every stage of the launcher. This result can be justified by the assumptions considered for wiring harness. Solution would likely change if more complex harness constraints were considered.

\begin{figure}[htbp]
\centerline{\includegraphics[width=\columnwidth]{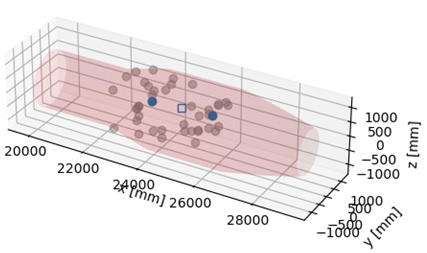}}
\caption{Visual representation of the optimal WSN topology for stage 5.}
\label{fig:stage5_topology}
\end{figure}

As an example, Fig.~\ref{fig:stage5_topology} shows graphically the results of the WSN topology optimization for stage 5, illustrating the positions of the sensors within the segment; grey points represent the sensors connected to the nearest sensor RF kit by wire, blue points represent the sensor RF kits, the empty blue square represents the HUB. Furthermore, Table~\ref{tab:stage5_hubs} reports the details of the HUB belonging to the stage 5, including the minimum transmitting power needed for matching the receiver sensitivity of all of the sensor RF kits. Instead, Table~\ref{tab:stage5_sensor_RF_kits} reports the details of the sensor RF kits belonging to the stage 5, including the label of the HUB each of them is paired with and the minimum transmitting power needed for matching the receiver sensitivity of the HUB.
Finally, the total emitted power by all of the transmitters in the wireless sensor network is 9.993141 dBm. 

\begin{table}[htbp]
\caption{HUBs within stage 5}
\begin{center}
\begin{tabular}{|c|c|c|c|c|}
\hline
\textbf{Id.} & \textbf{$X$ [mm]} & \textbf{$Y$ [mm]} & \textbf{$Z$ [mm]} & \textbf{Power [dBm]} \\
\hline
0 & 24469,5 & 432,75 & -9,15 & 6,675521 \\
\hline
\end{tabular}
\label{tab:stage5_hubs}
\end{center}
\end{table}

\begin{table}[htbp]
\caption{Sensor RF kits within stage 5}
\begin{center}
\begin{tabular}{|c|c|c|c|c|c|}
\hline
\textbf{Id.} & \textbf{$X$ [mm]} & \textbf{$Y$ [mm]} & \textbf{$Z$ [mm]} & \textbf{Power [dBm]} & \textbf{Hub} \\
\hline
0 & 25540 & 434,3 & 50 & 6,675521 & 0 \\
1 & 23399 & 431,2 & -68,3 & -1,66156 & 0 \\
\hline
\end{tabular}
\label{tab:stage5_sensor_RF_kits}
\end{center}
\end{table}

\section{Conclusion}

This work proposes a methodological framework for designing the optimal topology of WSNs in new generation launchers, and simulating the relevant features of the electromagnetic scenario determined by the topology itself. The framework consists of a reusable workflow accompanied by a reference toolset. For the specific case study of this work, the toolset is composed of Python scripts implementing the operations required by the workflow. The toolset has been implemented by using open-source libraries, except for the CEM evaluations which have been carried out using the commercial software suite AEDT.

In the case study, for each of the analysed segments of the launcher, the found optimal topology requires the deployment of only two sensor RF kits working as multiplexers for several sensor data sources, and one single HUB. This result can be justified by the assumptions considered for wiring harness; solution would likely change if more complex harness constraints are considered. The obtained results in terms of minimum power needed for reach the sensitivity of the receivers, allow to significantly contain both the energy consumption and the interference.

\textcolor{black}{Future work will firstly exploit analytical and statistical characterizations to overcome the limitations of the proposed framework. In details, as a methodological extension, we will investigate the computation of the inner scattering matrix by using analytical and/or semi-analytical approaches, e.g, based on tapered waveguides theory. Such scattering matrix will be used to directly optimize the network topology, taking into account also multipath and near-field effects. Instead, cable paths will be modelled by using a statistical framework in order to take into account both stage dimensions and the interior complexity. Such studies are crucial in order to obtain a general model, whose parameters must be linked to the synthetic specifications of the launcher as well as the communication system (e.g., dimensions, interior complexity, frequency, polarization).
Furthermore, the aforementioned methodological extension will be used to characterize the optimality of the proposed framework in this work, by providing a detailed study of the effects of the considered assumptions and design choices on the cost function behaviour.}

\textcolor{black}{Finally, the outcomes of the analysis carried on by means of the framework(s) will be used for the final configuration synthesis for (i) verifying if the electromagnetic environment induced by the wireless sensor network interferes with the avionics and other sensors (non-wireless) already installed in the stages, and (ii) defining possible optimized geometric layouts for further reducing interferences.}








\bibliographystyle{IEEEtran} 
\bibliography{biblio} 


\end{document}